# Distributed In-memory Data Management for Workflow Executions

Renan Souza[a, e, ✉], Vítor Silva[b, f], Alexandre A. B. Lima[a, b],
Daniel de Oliveira[c], Patrick Valduriez[d], Marta Mattoso[a, ✉]

[a] *COPPE/Federal University of Rio de Janeiro*
[b] *Campus Duque de Caxias / Federal University of Rio de Janeiro*
[c] *Fluminense Federal University, Brazil*
[d] *Inria, University of Montpellier, CNRS, LIRMM, France*
[e] Current affiliation: IBM Research, Brazil
[f] Current affiliation: Snap Inc., United States

**Abstract**

Complex scientific experiments from various domains are typically modeled as workflows and executed on large-scale machines using a Parallel Workflow Management System (WMS). Since such executions usually last for hours or days, some WMSs provide user steering support, *i.e.*, they allow users to run data analyses and, depending on the results, adapt the workflows at runtime. A challenge in the parallel execution control design is to manage workflow data for efficient executions while enabling user steering support. Data access for high scalability is typically transaction-oriented, while for data analysis, it is online analytical-oriented so that managing such hybrid workloads makes the challenge even harder. In this work, we present SchalaDB, an architecture with a set of design principles and techniques based on distributed in-memory data management for efficient workflow execution control and user steering. We propose a distributed data design for scalable workflow task scheduling and high availability driven by a parallel and distributed in-memory DBMS. To evaluate our proposal, we develop d-Chiron, a WMS designed according to SchalaDB's principles. We carry out an extensive experimental evaluation on an HPC cluster with up to 960 computing cores. Among other analyses, we show that even when running data analyses for user steering, SchalaDB's overhead is negligible for workloads composed of hundreds of concurrent tasks on shared data. Our results encourage workflow engine developers to follow a parallel and distributed data-oriented approach not only for scheduling and monitoring but also for user steering.

## 1. Introduction

With the evolution of computational tools and hardware, the ever-growing amount of data, and the increasing use of machine learning methods, more and more scientists from a wide variety of domains both in industry and academia have been using large-scale computers to conduct their experiments. A widely adopted strategy is to model the experiments as workflows and execute them using Parallel Workflow Management Systems (WMSs) on large-scale machines, such as High-Performance Computing (HPC) clusters [11]. Since these workflows typically execute for long hours or even days, users cannot wait until the end of the execution to start analyzing the workflow data and fine-tune parameters and convergence settings [2,11]. Despite solutions for self-tuning based on machine learning [40], decisions like changing convergence values, the number of iterations, or levels of interpolation still need human interference, which complements AI-based solutions [19,52]. Supporting user steering in scientific experiments allows users to run data analyses at runtime (*e.g.*, inspect, debug, visualize, monitor) that may lead to dynamic adaptation of aspects of the workflow (*e.g.*, change the input data, parameters, convergence criteria) [6,17,26,29,30].

In our past work with Chiron WMS [34,35], we have shown how users can have a better understanding of the data being processed in their experiments while the workflows are executing [42,43]. Then, based on such understanding, users may decide to dynamically adapt the workflow, such as reduce datasets [49], change user-defined loop conditions [13], and parameters [5,48]. Chiron adopts an integrated data management solution to store data dependencies, execution data, domain data, and provenance data all together in the same database available for monitoring and user steering. Suppose a parallel parameter sweep workflow with an *Activity 1* that uses a parameter $X$ to calculate a value $Y$, and an *Activity 2*, chained with the *Activity 1*, that delivers a result $Z$ for each input data of *Activity 1*. Chiron can answer queries like: "What is the current average value for parameter $X$ leading to the best $Z$ results?" or "List the status information about the 5 computing nodes with the greatest number of *Activity 1* tasks that are consuming input data that contain parameter $X$ values greater than 70." These are simple examples of analytical queries that can get overly complex as the user explores the data in an *ad-hoc* way, requiring several joins involving scheduling tables and different provenance tables. However, Chiron employs a centralized execution control to schedule tasks, which limits its scalability.

WMSs that are able to scale to thousands of CPU cores in HPC clusters, such as EMEWS (which runs on top of Swift/T) [24,53] and Pegasus [10,12], do not allow for user steering. Scalable workflow executions on HPC clusters require managing several types of data, *e.g.*, work queues, task data, performance data, provenance data, and other related data structures used by the workflow execution. Scheduling is fundamental for efficiency and is driven by work queues, table of events, execution status, and mapping data to tasks [1]. This is particularly critical in Many-Task Computing (MTC) [39] applications, where thousands of parallel tasks must be scheduled to multiple computing nodes, and each task consumes input data, performs computations, and produces vast amounts of data [39]. Since these features are efficiently supported by DBMSs, WMSs like Pegasus [10,12] have migrated their task queue scheduling management to a DBMS, but register data for user analyses in a separate and different DBMS. The EMEWS workflow system [24] makes data available for queries only after the workflow execution finishes and Pegasus has recently provided a user monitoring dashboard, but it is disconnected from the workflow engine data. This user data monitoring approach prevents the WMS from having the human in the loop of the workflow execution, which is essential for steering. Although user steering has been addressed for decades in computational steering environments, portals, and application-specific systems [33] it is still an open problem in the scientific workflow community [2,11,16,30]. To the best of our knowledge, there is no scalable workflow execution management approach capable of integrating, at runtime, execution, domain, and provenance data aiming at supporting user steering.

In this work, we propose a generic architecture with a set of design principles and techniques for integrating workflow scheduling data management with provenance and domain data to provide for an efficient user steering support. We call it SchalaDB: scalable workflow scheduling driven by an in-memory distributed DBMS. The data shared by scheduling and provenance is only written once, avoiding redundant operations and having less data movements. Also, the same DBMS cache can improve main memory access. Even though task queue data and provenance data are small in terms of size, their data management during the workflow execution is quite complex. The DBMS has to be efficient both for analytical queries and for concurrent update transactions. As discussed by Chavan *et al.* [7], the distributed in-memory DBMS approach is efficient even with mixed transactional and analytic workloads. A distributed data design is elaborated for providing data to scalable workflow task scheduling and high availability. Data extraction and ingestion at the DBMS allow for execution, provenance, and domain data to be available for runtime analytical queries, enabling users to adapt the workflow execution and improve its overall execution time based on steering.

To evaluate our approach, SchalaDB's design principles are implemented in Chiron WMS. Chiron's engine is modified by replacing its centralized task scheduler with a distributed one that obtains data through an in-memory DBMS. We call d-Chiron the resulting WMS that manages its data using SchalaDB principles [9]. In an extensive performance evaluation, we run both synthetic and real workflow workloads on an HPC cluster using up to 960 computing cores. Among other analyses, we show that SchalaDB's overhead is negligible for workloads composed of many concurrent long tasks, typical in scientific workflows. Our results encourage workflow system developers to follow a parallel and distributed data-oriented approach, such as SchalaDB, not only for scheduling and monitoring, but also for user steering support.

The remainder of this paper is organized as follows. In Section 2, we discuss the data that need to be managed during a scientific workflow execution. In Section 3, we present SchalaDB design, architecture, and techniques. In Section 4, we present d-Chiron as an implementation of SchalaDB. In Section 5, we show the experimental evaluation. In Section 6, we show related work and we conclude in Section 7.

## 2. Data Management in Large-scale Workflows

Data management in large-scale scientific workflow execution is a major challenge because it has to deal with several types of data. We can group these types of data into three categories: (i) execution, (ii) domain, and (iii) provenance. In this section, we briefly explain each of them. Although presented separately, we notice that those categories share a lot of data [34,36,42,43,49]. Storing them separately leads to data redundancy and lack of data integration support for runtime data analysis.

## 2.1 Execution Data

Task scheduling is a basic execution control functionality of any WMS. Other parallel execution control features, such as availability and concurrency are also very important. By providing efficient parallel access to execution data, there is less contention on scheduling data structures. More specifically, work queue is the main data structure for task scheduling in MTC, holding the list of tasks (following specific dependencies) to be scheduled among the computing nodes (*i.e.*, machine nodes composing an HPC cluster) [39]. Information such as which tasks should be scheduled to which computing nodes, number of tasks per node, tasks' input data, tasks' duration, and how much memory or computing power were consumed are examples of execution data to be managed in an MTC scheduler. Dedicated DBMSs have started to be used by execution engines to manage scheduling.

## 2.2 Provenance Data

Scientific workflows need provenance data, since they allow for analysis, quality assessment, reliability, reproducibility, and reusability of the scientific results [8,18]. Provenance data track which and how processes were utilized to derive each data entity; keep data authorship; and provide powerful means for analytical queries to perform data reduction, discovery and interpretation [28]. Such features are considered as important as the scientific achievement itself, since experiment's reliability can be compromised otherwise [18]. Provenance data representation is much more than registering what was executed. The W3C PROV [20] recommendation allows for a generic and uniform provenance data representation, which promotes interoperability and data analyses in general. In scientific workflows, provenance data received PROV specializations, like PROVOne [4] and PROV-DfA [47] to cover information about workflows' specifications, agents, activities and data derivation paths. Provenance data management requires capturing data, explicitly relating them to the workflow activities and efficiently storing these data to keep workflow high performance execution.

## 2.3 Domain Data

Although scientific domain data are typically managed by the simulation programs composing the workflow, the WMS has to be aware of some of them to manage the dataflow and register provenance data. Enabling analyses with domain data enhances analytical capabilities in a WMS. Defining how much domain data should be known by the WMS and represented in the provenance database is challenging [42].

Domain data management in scientific workflows is intrinsically hard. The scenario typically involves a high number of raw data files, multiple directories and subdirectories, and a wide variety of file formats, such as text-based (CSV files and plain-text matrices), and binary-based (like HDF5, FITS, NetCDF, and SEG-Y). Registering pointers to these large raw data files with some relevant raw data related to the dataflow is already a significant help in runtime workflow data analysis [42].

## 3. SchalaDB: Scalable Distributed Data Management for Workflow Executions

SchalaDB is a reference architecture that follows a set of techniques, based on distributed data management principles, for scalable user steering support of workflow executions. Distributed workflow execution control requires managing a large work queue with concurrent access as well as managing a variety of data about tasks, input and output domain values, and provenance. Capturing, structuring and loading these data in a DBMS for runtime data analyses is beneficial to users, but it may interfere on the workflow execution performance. Our main goal with SchalaDB is to allow users to steer the workflows, based on runtime analyses of its database, without harming the overall workflow performance when capturing data for analysis. SchalaDB innovates by exploiting an in-memory high-performance DBMS as an integrated data provider for workflow execution and user steering. The motivations to use such a DBMS in SchalaDB are as follows.

First, DBMSs already implement very efficient mechanisms that are essential in HPC, such as concurrency control and fault recovery. Particularly, guaranteeing Atomicity, Consistency, Isolation, and Durability (ACID) properties for update transactions is useful when task-related data management suffers multiple concurrent updates on the work queue during scheduling. DBMSs, especially those that are cluster-based, enable robust parallel cache memory usage. Furthermore, data replication and partitioning into multiple nodes are well studied and implemented in such DBMSs. A partitioned work queue is potentially faster to query and update than a centralized one. Inserting and removing tasks as well as querying parallel tasks' status are operations directly handled by the DBMS.

In the following sections, details about how SchalaDB combines high performance with powerful data analyses support are given. Section 3.1 presents SchalaDB architecture and shows how high performance data management techniques can be used considering different levels of parallelism between workflow execution and a DBMS; in section 3.2, techniques on how to partition and distribute data in our context – particularly the work queue – are described; and, section 3.3 shows how distributed data management can be used to implement execution control relying on the DBMS

### 3.1 SchalaDB Architecture and Techniques

Several execution control activities (*e.g.*, scheduling, and availability) must be carried out in the course of a workflow execution. After analyzing and comparing many WMS' architectures, Liu *et al.* [27] identify similarities between them and propose a generic WMS Functional Architecture. Figure 1 shows this architecture, which comprises five layers, each of them providing services to be used by the layer immediately above. One layer, named "Workflow Execution Plan (WEP) Execution", is dedicated to execution control activities or modules: "scheduling", "task execution" and "fault tolerance" [27]. Three layers lie above this one, showing how important these activities are to WMSs. Thus, techniques to improve them, like those proposed by SchalaDB, benefit WMSs in many ways.

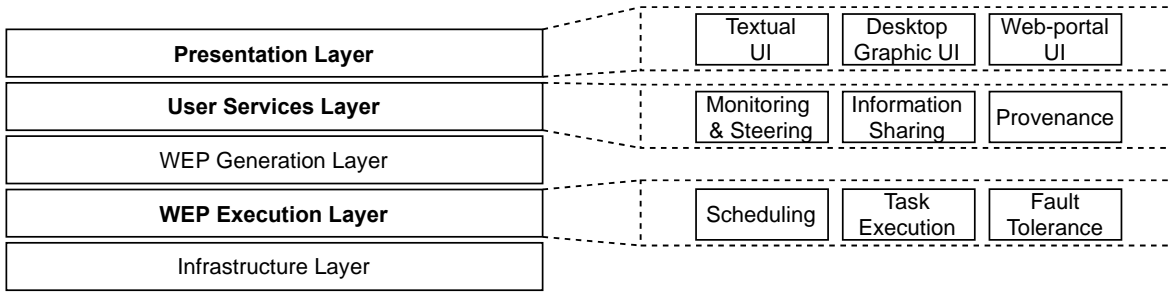

Figure 1. WMS Functional Architecture (adapted from Liu *et al.* [27]).

SchalaDB also directly contributes to another layer of the WMS Functional Architecture [27]: the "User Services" layer, more specifically to its "workflow monitoring and steering", and "provenance data" modules. Supporting these kinds of services is SchalaDB's main goal. The "Presentation" layer has modules responsible for interfacing between the end-users and the WMS. These are discussed in the next section, when describing the implementation of SchalaDB in a WMS.

In this section, we give an overview of SchalaDB architecture and how it uses a DBMS to explore workflow parallel execution control and different levels of parallelism.

In an in-memory DBMS, the nodes that run database operations are often called *data nodes*, accessing the distributed memory. Each *data node* typically contains multiple cores. The DBMS client nodes are called *worker nodes*. In a workflow execution, each *worker node* also contains multiple cores for parallel and distributed computing of the scientific workflow activities.

**SchalaDB Architecture.** In Figure 2, we present SchalaDB architecture. Components responsible for workflow execution control are organized as a multiple-master/multiple-worker nodes architecture. Instead of having multiple workflow *master* nodes that actively distribute tasks for *worker nodes*, SchalaDB adopts *data nodes* to manage data structures like the Work Queue (WQ) that are queried by *worker nodes* that demand tasks. In SchalaDB, WQ data is distributed across $D$ *data nodes*. Each *worker* $w_i$ executes workflow tasks (the actual scientific computation). The number of *data nodes* is typically much smaller than the number of *worker nodes*. The goal is to privilege workflow parallel execution rather than the database operations, which are much smaller. *Connectors* are brokers that intermediate the communication between the DBMS and other components. They are implemented using DBMS drivers. *Supervisor* is responsible for adding tasks to the WQ. *Secondary supervisor* eliminates the single point of failure by becoming the main supervisor in case the original main supervisor crashes.

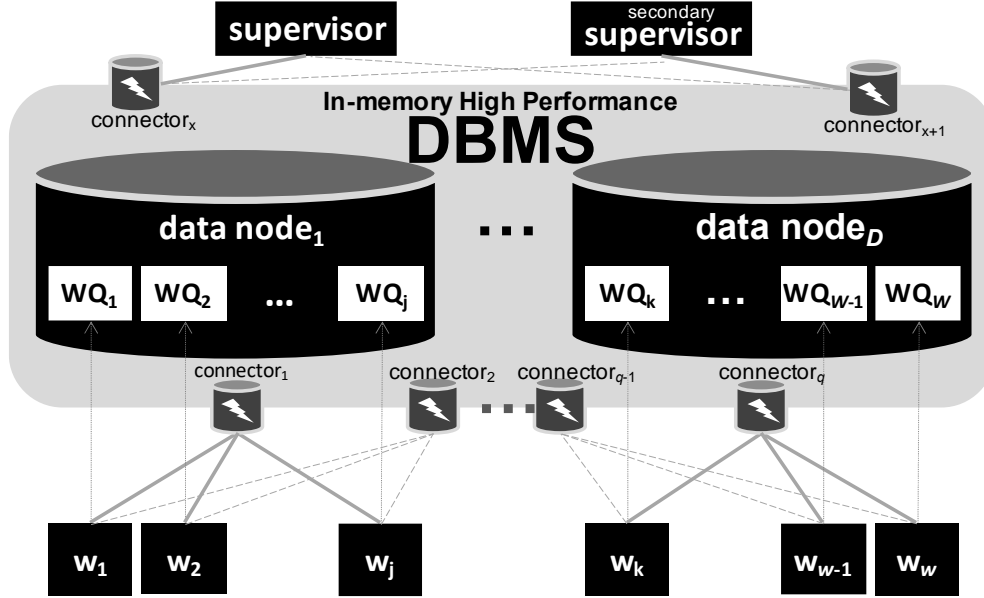

**Figure 2. SchalaDB architecture.** *W* worker nodes directly accessing the DBMS composed of *D* data nodes.

**Allocation flexibility**. Except for *supervisors* (one *supervisor*, one *secondary supervisor*), SchalaDB does not require any specific number of components of each type, as long as there are at least one of each. Also, more than one component may be allocated to the same physical node. While this gives a lot of flexibility for workflow execution, these allocation choices may impact performance. For higher availability and locality purposes, we recommend that each physical node hosts no more than one component of each kind. For example, one given physical node may run a *data* and a *worker node*s, but neither two *worker nodes* nor two *data nodes*.

**Scheduling.** Scheduling adopts a passive multi-master approach where *worker nodes* obtain tasks from WQ by querying *data nodes*. By using a DBMS, the capability of serving multiple concurrent requests from *worker nodes* is given. This approach differs from traditional master-worker WQ implementations where *worker nodes* request tasks from a *master* through regular message passing, such as MPI. By exploiting DBMS concurrency control, SchalaDB avoids the overhead of developing concurrency control algorithms. Therefore, having a distributed WQ (*i.e.*, no centralized execution data) reduces contention problems and improves workflow parallel execution performance. Even if workflow executions of certain experiments do not demand a distributed WQ, having a distributed in-memory DBMS to manage it would not negatively impact WMS performance. Besides, such a WMS would already be prepared for more demanding workloads.

**Availability.** Each SchalaDB component may be replicated for high availability. Regarding data managed by the DBMS, data replication techniques can be directly employed for *data nodes* automatic failure recovery. Considering *worker nodes*, each *worker* $w_i$ may improve availability by connecting and querying the DBMS via different *database connectors*. In Figure 2, full gray lines represent links between *worker nodes* and their respective main or primary *connectors* while dashed gray lines represent links between *worker nodes* and their backup or secondary *connectors*. If a *connector* fails, all *worker nodes* connected to it are switched to their secondary ones. Therefore, the *secondary supervisor* removes the single point of failure from the *supervisor* node.

***Worker nodes* to *connectors* distribution.** *Connectors* are usually responsible for just listening to connections from worker nodes. We propose a simple strategy for distributing *worker nodes* to *connectors*. First, if a *worker node* shares a physical node with a *connector*, then this is its primary connector. Then, remaining *worker nodes* are distributed to *connectors* by using a simple round-robin strategy.

**Adherence to WMS Functional Architecture.** As we can notice, SchalaDB architecture's components implement all services or modules described as part of the "WEP Execution" layer from WMS Functional Architecture [27]. Besides, having a WQ managed by a high-performance DBMS provides a good basis for the implementation of the "workflow monitoring and steering" and "provenance data" modules that belong to the "User Services" layer of the referred architecture. WQ data is part of provenance data. The DBMS makes it possible to perform complex queries and updates on such data at runtime, enabling workflow monitoring and steering.

## 3.2 SchalaDB Techniques for Data Partitioning

In this section, we explain SchalaDB design techniques for data partitioning. DBMSs typically have several partitioning techniques like round-robin, hashing, and by value ranges [38]. Considering the number of elements, WQ is typically the largest scheduler data structure. Without loss of generality, we assume that a Relational DBMS is used for managing workflow execution data. The same design techniques can be employed with other data models. This way, we consider WQ is implemented as a relation comprised by tuples representing tasks.

The first design step is about partitioning WQ data. Our proposal is to hash partition WQ based on the *worker identifier (id)* assigned to the task. Considering a SchalaDB architecture instance with $W$ *worker nodes*, WQ has $W$ partitions. The goal here is to initially produce partitions with similar size. The second design step deals with allocating WQ partitions, *i.e.*, assigning partitions to *data nodes*. Database processing is much lighter than workflow execution. Therefore, there are usually much more partitions than *data nodes*. Having more partitions than *data nodes* gives flexibility to implement more sophisticated load balancing techniques without data transfer between *data nodes* and avoids data transfer and communication overhead. The third design step is about defining replicas. Replicating database relations may be advantageous for fault tolerance and query processing. On the downside, *worker nodes* typically present highly concurrent transactions during scheduling, as tasks are very often updated when scheduled, executed and completed, which can be time consuming. Therefore, SchalaDB adopts one replica for each partition.

Despite having different *workers* accessing the same *data node*, data parallelism is improved when each *worker node* accesses its own WQ partition as different memory spaces can be accessed in parallel for each partition. Local processing is also improved because task lookup for each *worker node* will go straight to its partition instead of searching in a potentially large shared partition. This also reduces race conditions among different *worker nodes*, which otherwise would be competing for an entire WQ partition. Each *worker node* $w_i$ gets or modifies tasks within its $WQ_i$ partition by submitting queries like "select/update the next ready tasks in the WQ where $worker\_id = i$". Figure 3 shows an excerpt of a WQ relation with execution data when a synthetic workflow composed of three activities was running on a small cluster with two *worker* nodes, each with two cores, and one *data node*. Thence, two *worker nodes* were running, each one on a different computing node, and the single *data node* was managing the WQ with two partitions. This is just an exemplary illustration of the WQ partitioning with execution data.

| Task Id | Act Id | Worker Id | Core | Command Line | Work space | Fail. Trials | Std Out | Start Time | End Time | Status |
|---|---|---|---|---|---|---|---|---|---|---|
| 1 | 1 | 1 | 1 | ./run a=1.3 b=27.75 c=16.21 | /data/act1 | 0 | x=18.71 y=6.77 | 2017-06-04 09:55:04 | 2017-06-04 09:55:58 | FINISHED |
| 3 | 1 | 1 | 2 | ./run a=0.67 b=19.18 c=24.26 | /data/act1 | 0 | x=4.58 y=0.39 | 2017-06-04 09:55:04 | 2017-06-04 09:55:59 | FINISHED |
| 5 | 2 | 1 | 1 | ./run a=1.9 b=17.96 c=23.92 | /data/act2 | | | 2017-06-04 09:56:00 | | RUNNING |
| 7 | 2 | 1 | 2 | ./run a=2.73 b=35.74 c=24.55 | /data/act2 | 0 | x=1.74 y=7.17 | 2017-06-04 09:55:59 | 2017-06-04 09:56:13 | FINISHED |
| 9 | 3 | 1 | 1 | ./run a=0.55 b=29.48 c=16.66 | /data/act3 | | | | | READY |
| 11 | 3 | 1 | 2 | ./run a=2.6 b=30.1 c=13.66 | /data/act3 | | | 2017-06-04 09:56:13 | | RUNNING |
| 2 | 1 | 2 | 1 | ./run a=1.49 b=6.64 c=9.22 | /data/act1 | | | 2017-06-04 09:55:04 | | RUNNING |
| 4 | 1 | 2 | 2 | ./run a=0.17 b=30.65 c=12.61 | /data/act1 | 0 | x=8.08 y=8.5 | 2017-06-04 09:55:03 | 2017-06-04 09:56:04 | FINISHED |
| 6 | 2 | 2 | 1 | ./run a=0.54 b=23.45 c=24.57 | /data/act2 | | | | | READY |
| 8 | 2 | 2 | 2 | ./run a=2.2 b=13.87 c=19.84 | /data/act2 | | | 2017-06-04 09:56:05 | | RUNNING |
| 10 | 3 | 2 | 1 | ./run a=0.48 b=18.39 c=16.79 | /data/act3 | | | | | READY |
| 12 | 3 | 2 | 2 | ./run a=0.59 b=15.67 c=13.06 | /data/act3 | | | | | READY |

**Figure 3. Excerpt of a WQ relation with 2 partitions. Background colors represent WQ partitions.**

## 4. d-Chiron WMS: An Implementation of SchalaDB

Chiron WMS is open-source software and, to the best of our knowledge, it is the only WMS that manages execution, domain, and provenance data, jointly, using a DBMS (PostgreSQL) at runtime, thus enabling enhanced runtime data analyses through SQL. Chiron provides W3C PROV compliant provenance data. Since Chiron is already based on a DBMS, we opted for modifying its engine in order to produce, as a proof of concept, a functional implementation of SchalaDB's architecture and design principles. Such an effort resulted in d-Chiron, a distributed WMS that uses SchalaDB design principles. MySQL Cluster [37] was chosen as its DBMS because it is high performant in HPC, open source, in-memory, distributed, and performs well both for update transactions and for joining several tables in ad-hoc queries. This section explains what was modified in Chiron to adhere to SchalaDB, originating d-Chiron.

**Chiron and d-Chiron**. With respect to the WMS Functional Architecture (*c.f.* Figure 1, [27]), Chiron has some modules in the "Presentation" layer that facilitate its usage in HPC clusters. One module helps users dispatch the WMS workers in the HPC cluster through a command line interface (CLI), which is represented by "Textual UI" in the WMS Functional Architecture. Also, there is a CLI to ease running steering queries to the database. Another CLI in this layer helps users create domain data tables in the database. In the "WEP Execution" layer, Chiron has the "scheduling" and "task execution" modules that implement a centralized execution control with a master-worker scheduling design. In Chiron, only the master node is able to access the centralized DBMS and to send tasks to the workers using MPI. In d-Chiron, to implement SchalaDB, we did major modifications in the "WEP Execution" layer of Chiron. More specifically, the execution control and task scheduling in d-Chiron is distributed as the workers have now direct access to the DBMS, using SQL queries, without needing to hop through a master. d-Chiron also implements the supervisor and secondary supervisor nodes of SchalaDB. Chiron and d-Chiron architectures are shown in Figure 4 and Figure 5, respectively.

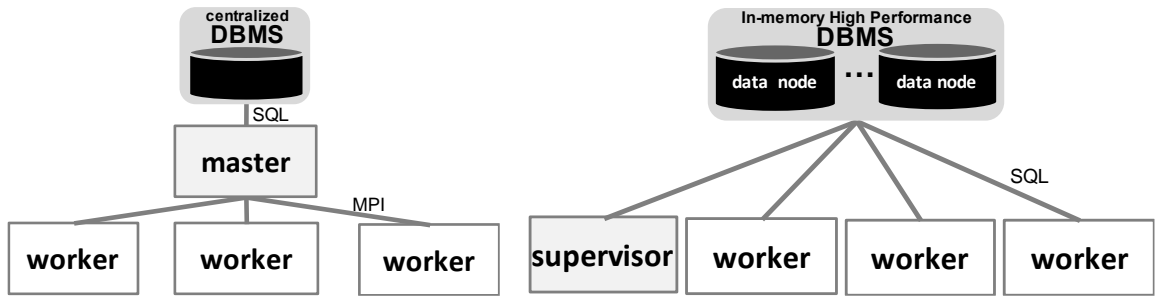

**Figure 4. Chiron centralized execution control.**    **Figure 5. d-Chiron distributed control driven by the DBMS.**

In Figure 6, we depict the differences between an MPI-driven master-worker scheduling, typically found in many scheduling systems in HPC solutions, and a distributed and parallel DBMS-driven distributed scheduling. In the later, there are less "proxies" between a *worker* and their tasks, as shown by the numbered items in Figure 6: each item is an operation and its respective number, the order of execution. In d-Chiron (Figure 6-A), a worker just needs to query the DBMS to get its tasks, update them, and store results. Using a centralized architecture (Figure 6-B), as in Chiron, since the centralized DBMS struggles to handle multiple parallel requests, there is the need for a master node to alleviate a severe bottleneck at the centralized DBMS. In the centralized architecture, the worker requests are first queued at the master, which submits queries to the centralized DBMS according to this auxiliary queue. In the centralized design, an additional acknowledgement message is needed so the worker can inform the master about tasks' executions. In summary, because of SchalaDB, d-Chiron's workers have direct access to their tasks, whereas, in Chiron, the workers need to ask a centralized master for tasks. Figure 6-A and Figure 6-B show that scheduling centralization leads to an increase in the number of processing steps; messages exchanged; complexity of implementation code regarding scheduling; and to potential bottlenecks during workflow execution.

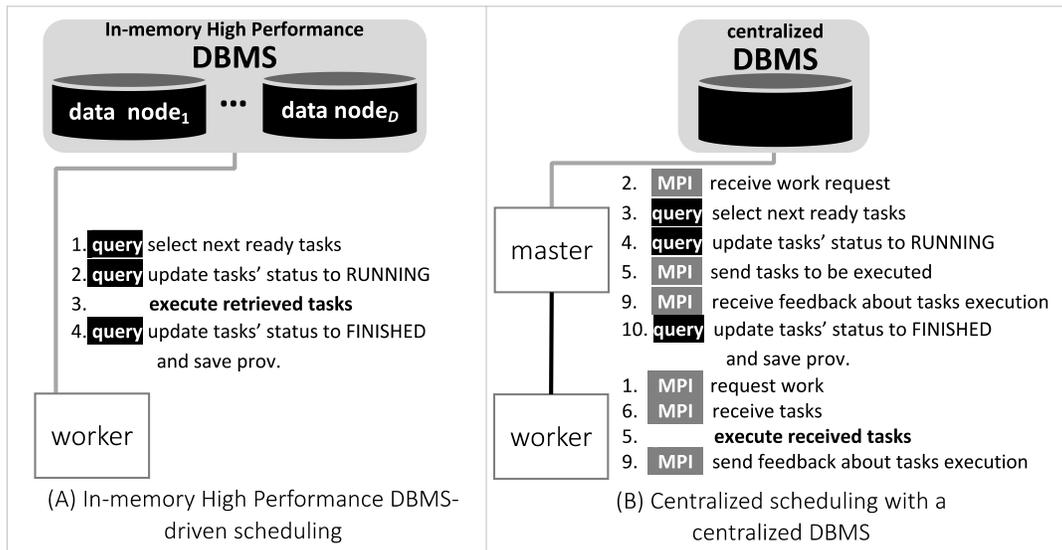

Figure 6. Comparison between centralized scheduling vs. DBMS-driven scheduling.

In Figure 6, we depict the differences between an MPI-driven master-worker scheduling, typically found in many scheduling systems in HPC solutions, and a distributed and parallel DBMS-driven distributed scheduling. In the later, there are less "proxies" between a *worker* and their tasks, as shown by the numbered items in Figure 6: each item is an operation and its respective number, the order of execution. In d-Chiron (Figure 6-A), a worker just needs to query the DBMS to get its tasks, update them, and store results. Using a centralized architecture (Figure 6-B), as in Chiron, since the centralized DBMS struggles to handle multiple parallel requests, there is the need for a master node to alleviate a severe bottleneck at the centralized DBMS. In the centralized architecture, the worker requests are first queued at the master, which submits queries to the centralized DBMS according to this auxiliary queue. In the centralized design, an additional acknowledgement message is needed so the worker can inform the master about tasks' executions. In summary, because of SchalaDB, d-Chiron's workers have direct access to their tasks, whereas, in Chiron, the workers need to ask a centralized master for tasks. Figure 6-A and Figure 6-B show that scheduling centralization leads to an increase in the number of processing steps; messages exchanged; complexity of implementation code regarding scheduling; and to potential bottlenecks during workflow execution.

**d-Chiron DBManager**. Instantiating a DBMS in multiple computing nodes is time consuming and error-prone, even for an experienced user. For this reason, we developed d-Chiron *DBManager*. By using this component, the WMS designer simply needs to adjust installation parameters, like how many of each DBMS components should be instantiated and on which computing node each of them should run. *DBManager* automatically instantiates MySQL Cluster, running its components in the appropriate computing nodes. We implemented a CLI for the *DBManager* component. An exemplary sequence of steps for running a workflow from scratch on an HPC cluster in d-Chiron is shown in Figure 7. First, *DBManager* initializes all DBMS components in preconfigured computing nodes (line 1). A user needs to run line 1 only once to initialize the WMS and DBMS processes in the cluster.

Then, the database is created (line 2). After that, workflow execution is started (line 3). During workflow execution, the user may run steering queries, as illustrated by line 4. After workflow completion and user's analyses, the user may decide to leave the DBMS up for future runs or, simply execute a shutdown command (line 5) to safely shut down the DBMS.

1. $> ./DChironDBManager --start
2. $> ./DChironSetup --create database
3. $> ./DChironStarter --start
4. $> ./DChironQueryProcessor --q "select * from workqueue where status = 'RUNNING' order by starttime"
5. $> ./DChironDBManager --shutdown

**Figure 7. Typical steps for running a workflow with d-Chiron.**

**Data Partitioning in d-Chiron.** Here we explain how we use MySQL Cluster to implement SchalaDB's data partitioning techniques in d-Chiron. Following SchalaDB techniques, the number of partitions is set to $W$, the number of worker nodes. However, in MTC, several tasks are managed by the same work queue partition. To reduce the impact caused by concurrent access to a shared partition, the supervisor circularly assigns a worker id to each task. Then, each worker $w_i$ gets or modifies tasks using queries with a selection predicate "$where\ worker\_id\ =\ i$".

Additionally, data allocation is delegated to MySQL Cluster as it efficiently balances the number of partitions per data nodes. We configure MySQL Cluster to replicate all relations across computing nodes running with replication factor set to one, meaning that each relation has one replica.

Finally, d-Chiron keeps the runtime analytical support previously available on Chiron. Domain, execution, and provenance data are still being captured and managed, but in a much more efficient and fault tolerant way.

## 5. Experimental Evaluation

In order to assess the benefits of SchalaDB, we perform a series of experiments with d-Chiron on an HPC cluster with up to 960 cores. This section describes the results obtained and is organized as follows. In Section 5.1, we present the experimental setup. In Section 5.2, we analyze several workloads to understand the impact of two dimensions in MTC workflows' scheduling: number of tasks and task duration. We evaluate long running tasks typical of scientific workflows as well as short running tasks. In Section 5.3, we analyze overhead incurred by SchalaDB. Finally, in Section 5.4, we show that our implementation of distributed in-memory data management in SchalaDB outperforms the implementation of a scheduler that relies on a centralized approach for the management of data for scheduling of parallel tasks by two orders of magnitude.

*5.1 Experimental Setup*

This section describes the workflow, workload, software and hardware resources used for tests.

**Workflow case study.** Risers Fatigue Analysis Workflow [49] is a real case study from the Oil & Gas industry. This workflow calculates the fatigue of ultra-deep oil platform structures, such as risers. Input data that represent environmental conditions (*e.g.*, as wind speed and wave frequency) are combined to evaluate stress on the riser's curvatures using seven linked workflow activities (Figure 8). As it takes a long time to calculate the fatigue for each environmental condition, users have to steer the execution so that some parameter ranges may be pruned out of the execution [49]. This parameter tuning depends on several data analyses and cannot be pre-programed for automatic pruning. In this workflow, computational engineers know how to fine tune input parameters values based on the specific behavior of the current workload. That is, the number of tasks the workflow is expected to run and how long each of them will take to perform its part of the application computation (*i.e.*, part of the processing that is exclusively related to the application behavior rather than processing related to workflow execution control). For this reason, this workflow is a good use case for our experiments, as we can vary such parameters to generate synthetic workloads we need for the tests.

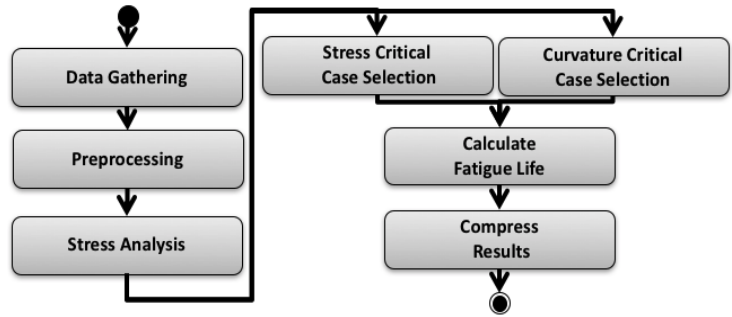

Figure 8. Risers Fatigue Analysis workflow.

**Workloads.** Based on the Risers workflow specification we generated several synthetic workloads with different combinations for the number of tasks and duration for the workflow activities. We repeat the experiments until the standard deviation of workflow elapsed times are less than 1%. The results are the average within the 1% margin.

**Software.** We use Chiron and d-Chiron. Executables for both systems can be found in the GitHub repository [9]. Chiron uses Postgres 9.5.1 and d-Chiron uses MySQL Cluster 7.4.9.

**Hardware**. The experiments are conducted on Grid5000 (www.grid5000.fr) using the StRemi cluster. Hardware specification is described in Table 1.

Table 1. Hardware specification of the HPC cluster in Grid5000.

| #Nodes | #Cores per node | Total cores | RAM per node | Processors | Network | Storage |
|---|---|---|---|---|---|---|
| 42 | 24 | **960** | 48 GB | AMD Opteron 6164 HE 1.7 GHz/12MB | Gigabit Ethernet | SATA AHCI & RAID-5 |

**Component-to-node allocation.** During the experiments, unless otherwise specified, d-Chiron components are allocated to the computing nodes in the cluster as follows. Each computing node runs a d-Chiron worker. Besides, in one of the computing nodes in the cluster, a supervisor runs alongside with a worker; and, in another node, a secondary supervisor runs alongside with a worker. Two SchalaDB's data nodes run on two other computing nodes in the cluster. Since the database usually has small sizes even for large workloads, as it stores only workflow's metadata (preliminary experiments [44,50] show

SchalaDB's database with tens of MB for large workloads), we opt for using only two nodes for running in-memory data nodes with occasional on-disk checkpoints, which is the minimum to achieve fault-tolerance. We set the maximum number of d-Chiron threads to be equal to the number of cores of each computing node.

*5.2 Scalability Analysis*

Scalability refers to the system's ability to deal with a growth of either the computing nodes (*e.g.*, addition of nodes) or the workload (*e.g.*, adding more data or tasks) [23]. Varying the number of computing nodes is straightforward. With respect to workload variation in MTC workflows, workloads are composed of thousands of tasks that must run in parallel and each one may last for seconds, minutes, or even hours [39]. Usually, the longer the task, the more complex it is, requiring complex scientific computations or extensive data manipulations.

We assess three types of scalability analysis: (a) Strong scaling (also known as speedup test), which intends to analyze how workflow execution time varies when we increase the number of computing nodes, but maintain the same workload. Ideally, the performance increases proportionally to the amount of computing nodes added. (b) Weak scaling, which intends to analyze how workflow execution time varies when we proportionally increase both the number of computing nodes and the workload, so that workload per processor remains constant. Ideally, the execution time should remain constant. (c) Workload scalability, which refers to analyzing how workflow execution time varies when we vary the workload but maintain the number of computing nodes fixed. Ideally, performance's variation should be proportional to workload variation. Since we consider a workload as composed of two factors (task duration and number of tasks), we can further fix one of them and increase the other in order to give finer understanding about the system's performance.

In the scalability analyses performed in this section, we set a base execution time for comparisons. When the workload is small, it is possible to execute the system running on a single CPU core and use this time as the base. However, most of the workloads used in this section would take weeks or months to completely finish on a single core, prohibiting this kind of execution. For this reason, in each experiment in this section we specify the base result, *e.g.*, when the least amount of parallel processes was used, and plot a linear execution time according to this base time.

**Experiment 1: Strong scaling with variation on the number of threads per process.** In this experiment, we investigate d-Chiron's strong scalability and the impact caused by concurrency on a larger cluster. Four cluster setups were employed: 120, 240, 480, and 960 cores (5, 10, 20 and 40 nodes, respectively). In order to investigate concurrency, for each setup we executed experiments varying the number of threads used by each worker: 12, 24 and 48 threads. Base execution time is set to the smallest number of cores evaluated, *i.e.*, 120 cores. We used a workload with 13 thousand tasks with mean task duration of 1 minute each. The results are in Figure 9(a), where we plot six curves: one for each setting of number of threads per process, along with calculated execution time for each setting if linear speedup was achieved. For simplicity, we refer to the latter curves as "linear time" or "linear executions", according to the context.

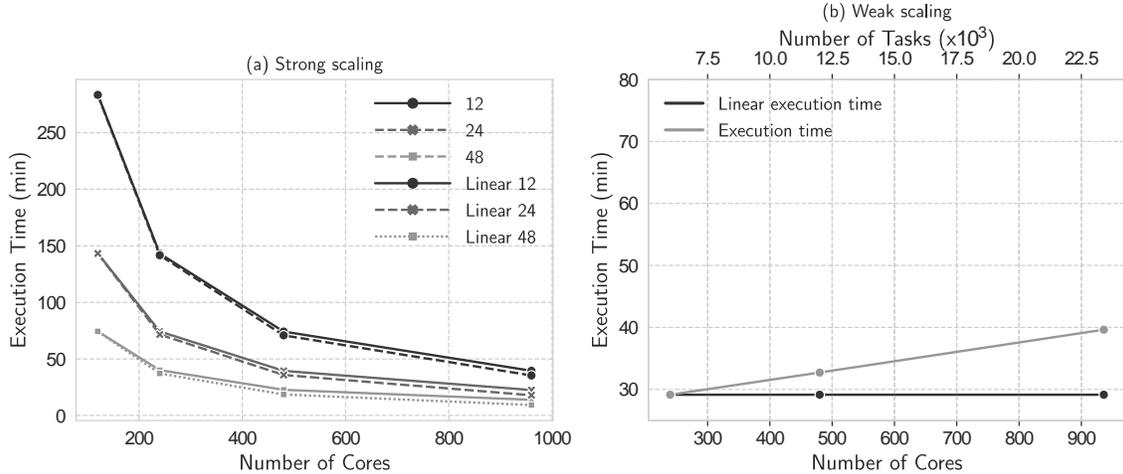

**Figure 9. Scaling analyses: (a) strong scaling and (b) weak scaling.**

Figure 9(a) shows that d-Chiron speedup attains close to linear in almost all cases. d-Chiron's speedup was almost linear with 12 and 24 threads per process for all configurations of nodes. However, it started to degrade speedup in the configuration with 48 threads per process and 40 allocated nodes (i.e., 960 cores in x-axis) since the number of threads was the double of the number of available cores (1920 SchalaDB workers). Despite of this speedup degradation, strong scalability experiment provides an important result, as we can use all CPU cores of a large cluster to perform the actual scientific computation of the managed application and d-Chiron still maintains high scalability.

**Experiment 2: Weak scaling.** Now we analyze how d-Chiron performs when we add computing nodes as we add more tasks to the workload. More specifically, we measure the execution time on 10 computing nodes (240 cores), 20 nodes (480 cores), and 39 nodes (936 cores) running about six, 12, and 23.4 thousand tasks, respectively. The mean task duration is one minute in these workloads. Figure 9(b) shows the results, where the base time is for the smallest number of cores tested (*i.e.*, 240) and the linear time means that by doubling the workload and number of computing nodes, ideally the execution time should remain constant. We use 24 threads per processor since it provided the best performance versus efficient usage of the available computational resources in the Experiment 1.

d-Chiron succeeds with weak scalability, as its curve remains close to the linear line. With 240 cores executing about 6 thousand tasks, the workflow finishes in 29 minutes. In 480 cores executing 12 thousand tasks, it takes 32.7 min. Ideally, it should also have taken about 29 minutes, *i.e.*, 3.5 min (or 12%) longer than the if linear scalability was achieved. With 936 cores nodes running about 23.4 thousand tasks, it takes 39 min, *i.e.*, 10.4 min (or 35%) longer than the linear time. Considering that d-Chiron is running up to almost 24 thousand tasks, from which 936 tasks are running in parallel and distributed among 39 computing nodes with 24 cores each, and that the linear time ignores intrinsic parallel management overhead, we find that these results are satisfying. Moreover, d-Chiron maintains a DBMS with online data for user queries and monitoring, which also introduces overhead, despite their advantages. This overhead is the subject of another experiment.

These results lead us to further investigate d-Chiron's performance by exploring a wider variety of workloads. Even though in MTC scientific workflows, tasks are considered to be long-term, *i.e.*, they may take many seconds or few minutes on average each [39], we now evaluate d-Chiron on a wider variety of workloads, including short-term tasks. The objective is to analyze d-Chiron's performance when we scale the workload in two dimensions: number of tasks and task duration. We run two experiments on 39 computing nodes (936 cores): fixed task durations, varying number of tasks; and fixed number of tasks, varying task duration.

**Experiment 3: Workload scalability – fixed task duration, varying number of tasks.** In this experiment, we analyze how d-Chiron performs when we vary number of tasks: from small (about 4.6 thousand), to mid (about 12 thousand), and large (about 23.4 thousand); and fixing two different task durations: short tasks (mean task duration of 5 seconds) and long tasks (mean task duration of 60 seconds). Similarly to the previous experiments, Figure 10(a) shows linear time for each setting; in this case, for each task duration (5s and 60s). We set the base case for the linear time to be the smallest number of tasks tested.

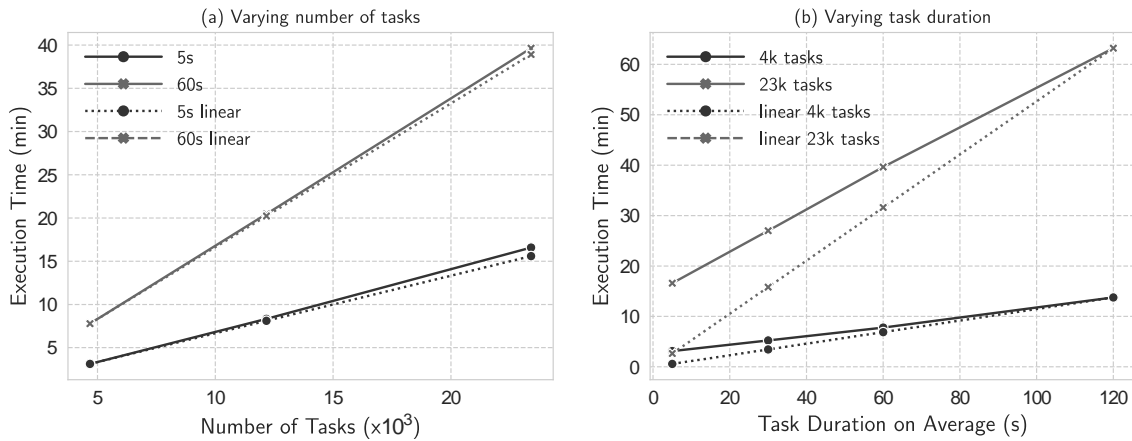

**Figure 10. Workload scalability analysis when varying (a) number of tasks and (b) task duration on average.**

In this experiment, d-Chiron attains close-to-linear performance for both task durations, for all number of tasks tested. For 5s, d-Chiron performs 2.7% and 6.3% worse than linear for 12 thousand and 23.4 thousand tasks, respectively. For 60s, d-Chiron performs 1.1% and 1.88% away from linear for 12 thousand and 23.4 thousand tasks, respectively. Therefore, by analyzing the increase in number of tasks, we see that d-Chiron's performance tends to decrease when the number of tasks increases, for both task durations (5 seconds and 60 seconds). More tasks mean larger WQ and more parallel management overhead. Also, the performance loss is slighter for the workload composed of longer tasks (60 seconds) than for short tasks (5 seconds). This result indicates that for longer tasks, d-Chiron attains higher scalability. We analyze this in more details in the next experiment.

**Experiment 4: Workload scaling – varying task duration, fixed number of tasks.** In this experiment, we analyze task duration variation more deeply to see how d-Chiron performs when we vary mean task duration from short (5 seconds as mean task duration) to longer (120 seconds), fixing two different number of tasks: small (4.6 thousand) and large (23.4 thousand, about 5 times larger).

We plot the linear line by setting the base result (where d-Chiron achieves best performance) as the longest task durations evaluated in this experiment, *i.e.*, the workload with mean task duration of 120 seconds. So, for instance, for a fixed number of tasks, if d-Chiron takes $T$ minutes to execute the workload composed of 120-seconds tasks, it should ideally take approximately $120/60 * T$ minutes to execute the workload composed of 60-seconds tasks. Figure 10(b) shows two linear curves, one for each fixed number of tasks evaluated (4.6 thousand and 23.4 thousand).

Analyzing the curves for 4.6 thousand tasks, we see d-Chiron runs close to the linear curve. We see that the longer each task takes, the closer to linear d-Chiron's performance tends to be. For the worst result, *i.e.*, for very short-term tasks, d-Chiron runs 4.6 thousand tasks in 3.1 minutes, whereas the linear would be in 0.6 minutes. Comparing the lines for 23.4 thousand tasks with the lines for 4.6 thousand, we see that d-Chiron is farther from the linear line. The worst result also occurs for the workload of 5-second tasks, but for longer tasks, d-Chiron performs well, even for the higher number of tasks. For example, for 60-second tasks, d-Chiron runs in 39.6 minutes, whereas the linear would be in 32 min (about 20% away from linear).

**Key findings:** From Experiments 3 and 4, we can conclude that our implementation of SchalaDB performs better when tasks take longer, *i.e.*, when they are more complex, which is the expected workload for real scientific workflows as discussed in Section 5.1. Also, as the number of tasks increases, the performance difference between short and long tasks becomes more significant. The reason for this is that too many small parallel tasks overload WQ, introducing large parallel management overhead and jeopardizing task scheduler's performance. In d-Chiron's case, most of this overhead is caused by an excessive amount of accesses to the DBMS in order to retrieve tasks from WQ. However, SchalaDB was designed targeting at scientific workflows, where real workloads are typically composed of long-lasting complex tasks, making runtime analysis a requirement.

*5.3 Assessing DBMS Impact on Performance*

In this section we measure every single DBMS access and see how they compare with the overall workflow execution time. Regarding software and hardware settings used for all experiments in this section, we use 39 computing nodes (936 CPU cores) in StRemi cluster.

**Experiment 5: Analysis of time spent accessing the DBMS.** In this experiment, we want to further understand the details of the overhead caused by DBMS accesses during workflow execution. In addition to the number of tasks, we investigate how much the spacing between multiple parallel accesses to the database (mimicked by the task duration) affects the performance, and how much of the total workflow time was dominated by DBMS accesses. We measure the elapsed time of every single query on the database made by each node at runtime. Then, for each node, we add up all elapsed times. As each node executes in parallel, we consider the time spent accessing the DBMS in a workflow execution as the maximum sum obtained this way.

We execute eight workloads composed of 23.4 thousand tasks each, with the following mean task durations, in seconds: 1, 2, 3, 4, 5, 10, 30 and 60s. Each workload represents different DBMS access patterns, varying from frequent concurrent accesses from many requesting worker nodes (1 second tasks) to more sparse accesses (60 seconds tasks). The results are in Figure 11, where the black bar represents the time spent in the DBMS and the gray bar represents the total workflow time, which include the time spent by the actual application being managed by the WMS and some other times – not related to DBMS accesses – spent for workflow execution management. These DBMS accesses occur in parallel and in background with the other operations.

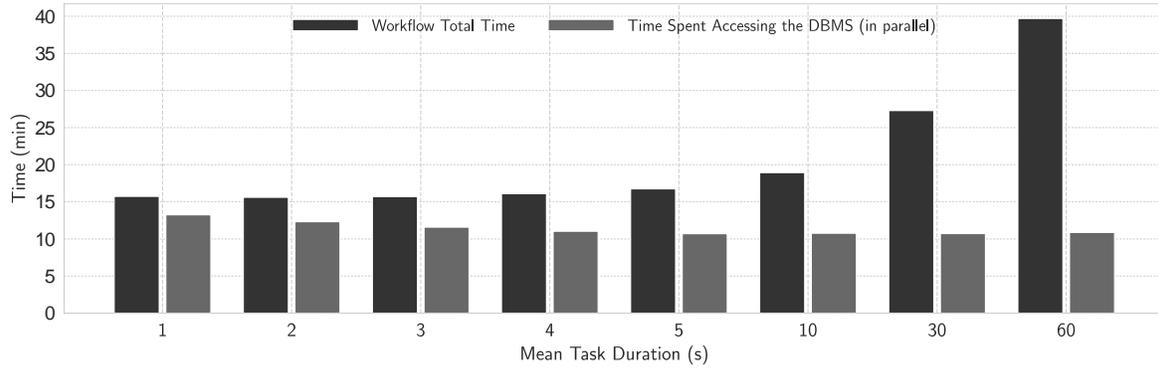

Figure 11. Analyzing impact of DBMS accesses in d-Chiron.

From Figure 11, we can identify a pattern for workloads where concurrent accesses occur more frequently (more evident from 1 to 3 seconds): the more frequent the concurrent DBMS accesses are, the greater the time spent accessing the DBMS. For such frequent accesses in a significant amount of tasks (23.4 thousand) and cores (936), the DBMS struggles to deal with so many frequent concurrent transactions, making it a bottleneck. This experiment also reveals that for workloads with such frequent accesses the DBMS access time is very close to the overall execution time. In other words, the overall execution time was almost completely dominated by time spent doing DBMS accesses. This result is important because we see that when the workload is composed of tasks that take at least 5 seconds, the DBMS access times do not depend on task duration. Thus, the negative impact of DBMS accesses is more significant for workloads composed of short-term tasks because the WMS takes more time performing DBMS accesses than application computation. However, when the workflow execution time is at least greater than two times the time spent with DBMS accesses, which happens for tasks that take about 25 seconds on average, the overhead introduced by the DBMS-based scheduler solution pays off, as the DBMS is not overwhelmed by so many concurrent transactions. Since DBMS operations run in background, for workloads dominated by tasks with mean duration of 25 seconds, d-Chiron is highly scalable as the DBMS overhead starts to become negligible when compared to application computation time.

Therefore, when the application computation time is greater than the time spent in accessing the DBMS, the overhead is amortized. This is the scenario where d-Chiron achieves best performance and is the case for most scientific applications, which often process workloads composed of tasks that take longer than one minute.

**Experiment 6: Describing d-Chiron's accesses to the DBMS.** In this experiment we analyze the time spent by main DBMS accesses at runtime by the workers. The results are in percentages with respect to the time spent by all DBMS accesses. We use the same workloads presented in Experiment 5. Since the number of DBMS accesses is proportional to the number of tasks in all workloads from Experiment 5 (*i.e.*, 23.4 thousand tasks), in this Experiment 6 we only present results for the workload with tasks with mean duration of 10 seconds. In Figure 12, we show percentages of time spent by each kind of DBMS access, relatively to the total time spent by all DBMS accesses. The total time is obtained as in Experiment 5, for the 10 seconds workload.

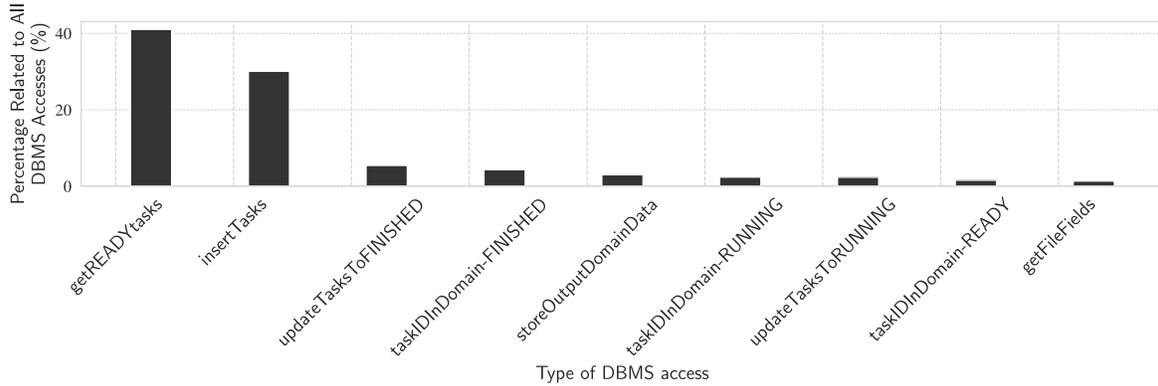

Figure 12. Analyzing DBMS accesses through specific SQL queries.

These accesses are all related to task scheduling. Tasks need to be inserted in the WQ as "READY" tasks. Then, each *worker* retrieves ready tasks, updates their status to "RUNNING", executes them, and finally updates their status to "FINISH". We can see that *getREADYtasks* by itself accounts for more than 40% of all DBMS accesses. Combined with *getFileFields*, these two operations represent 44.7% of read-only time spent accessing the DBMS. The other seven functions in Figure 12 are update operations and they account for 53% of the time spent accessing the DBMS. The remaining 2.3% are distributed over several shorter DBMS operations, both reads and writes. These are all update transaction-like query patterns (reading or writing specific rows of the WQ relation), confirming the main query pattern for task execution in d-Chiron, even though interactive analytical queries are also executed by users during workflow execution, as we show in Experiment 7.

These results also help to understand why DBMS access time is not only sensitive to mean task duration, but also to the number of tasks. The greater the number of tasks, the greater the number of those concurrent queries that need to update specific parts of the WQ relation. Each update makes the DBMS deal with distributed concurrency control, which is a complex operation. When these queries are not so frequent (*i.e.*, for longer task duration), concurrency is not so severe, making d-Chiron to achieve better performance.

**Experiment 7: Running user analytical queries at runtime.** Most of the experiments so far tested d-Chiron performance to understand how the DBMS-driven scheduling behaves under different workloads. In this experiment, we want to validate that the performance enhancements obtained after modifying Chiron according to SchalaDB principles, originating d-Chiron, do not affect system's ability to support user steering through runtime analytical queries. The support is maintained because d-Chiron still manages runtime provenance, execution, and domain data in its database. Data schema in d-Chiron is quite the same as in Chiron, it remains W3C PROV compliant. Therefore, in this experiment, we confirm that user steering support is maintained. Also, we intend to analyze performance overhead added to the workflow execution time by running such queries. For this, we run 8 typical user analytical queries in the Risers Fatigue Analysis domain during workflow execution. Their natural language descriptions are in Table 2 and their corresponding SQL code are available on GitHub [9], where we also provide the implemented data schema.

**Table 2. Analytical queries executed at runtime.**

| Q1 | Considering just tasks that started from one minute ago to now, determine tasks status, number of tasks that started, finished, and the total number of failure tries ordered by node. |
|---|---|
| Q2 | Given a node hostname, for each task, determine task status and the total size in bytes of the files consumed by the tasks that finished in the last minute. Order the results in descending order by bytes and ascending order by task status. |
| Q3 | Determine the hostname(s) of the nodes with the greatest number of tasks aborted or finished with errors in the last minute. |
| Q4 | Given a workflow identification, show how many tasks are left to be executed. |
| Q5 | Considering workflows that are running for more than one minute, determine the name(s) of the activity(ies) with the greatest number of unfinished tasks so far. Also, show the amount of such tasks. |
| Q6 | Determine the average and maximum execution times of tasks finished for each activity not finished. Show the name of the activity and order by average and maximum time descending. |
| Q7 | List cx, cy, cz, and raw data file path (output parameters produced in Pre-Processing activity) only when Calculate Wear and Tear activity produces f1 value greater than 0.5 and when the average time for the tasks in Calculate Wear and Tear activity takes more than average to finish. |
| Q8 | Based on a previous runtime analysis, modify input values to be consumed by the Analyze Risers activity, i.e., modify the input data for the next ready tasks for Analyze Risers activity. |

Queries from Q1 to Q6 analyze execution metadata and are of high importance for debugging and execution profiling. Q7 uses execution and provenance data, being an example of a query that relates dataflow output data of the fourth activity with output data of the second one, associated to deviations in execution time. Finally, Q8 represents a user that has analyzed Q7 results, made a decision, and then adapts the workflow by modifying input data for an intermediate workflow activity at runtime, illustrating d-Chiron's user steering capabilities.

We execute the workflow using a previously tested workload: 23.4 thousand tasks with mean task duration of 5 seconds each. We choose this workload because, as assessed in the previous experiments, workloads dominated by short-term tasks (such as 5-seconds lasting tasks) are more susceptible to higher latencies in d-Chiron due to a higher concurrent scenario. Thus, it is expected that if the overhead caused by the steering runtime queries is not high for this adversarial scenario, the overhead would remain low in a more favorable scenario with less concurrency. We conduct the experiment as follows. Using the 936 cores of the cluster, we run the workflow first without executing the set of steering queries, and then running each query in intervals of 15s during workflow execution. The results are in Figure 13.

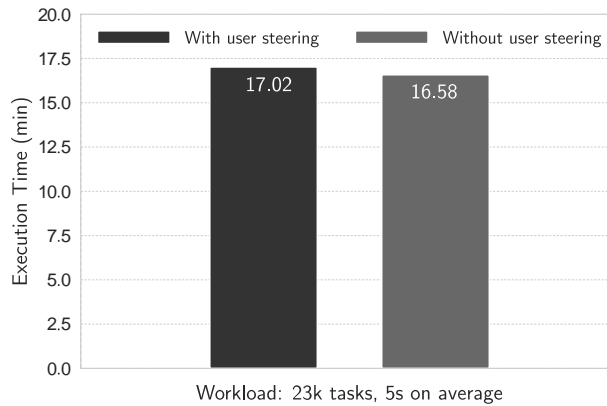

**Figure 13. Comparing overhead with and without runtime queries.**

Results show that the workflow execution time is approximately the same (less than 5% of difference between the scenarios with and without queries) no matter if running or not the queries at each 15 seconds, meaning that the overhead they cause is negligible. This happens because the database is managed by the DBMS and it is mostly composed of metadata rather than big raw data, which reside in files on disk, as discussed in the last experiment. Thus, queries run very fast (in the order of hundreds of milliseconds each) due to the reduced amount of data stored and parallel query processing in the in-memory DBMS. Additionally, in a typical scenario, there are not multiple scientists simultaneously monitoring the execution of a same workflow. This way, analytical queries are not expected to be executed very often. Even in situations where dashboards are employed for monitoring workflows, issuing queries more frequently (e.g., with refreshing intervals of one second), there is a small impact in workflows' execution times. In a previous work [49], we performed an experiment that submitted 30 parallel queries at every second against a d-Chiron's database and observed an overhead of 3.19% in an execution that lasted 17 min.

*5.4 Centralized vs. Distributed Execution Control and Task Scheduling*

Finally, the last experiment aims at evaluating the impact of decentralization on the scheduling and management of parallel tasks execution, our main goal when proposing SchalaDB. For this, we compare d-Chiron with Chiron.

**Experiment 8: Chiron vs. d-Chiron.** We evaluate four typical combinations of workloads of the risers workflow: (a) medium number of short tasks – 5 thousand tasks with mean task duration of 1 second each; (b) medium number of long-term tasks – 5 thousand tasks with mean task duration of 16 seconds; (c) large number of short tasks – 20 thousand tasks with mean task duration of 1 second; and (d) large number of long-term tasks – 20 thousand tasks with mean task duration of 16 seconds. Figure 14 shows the execution times with 936 cores.

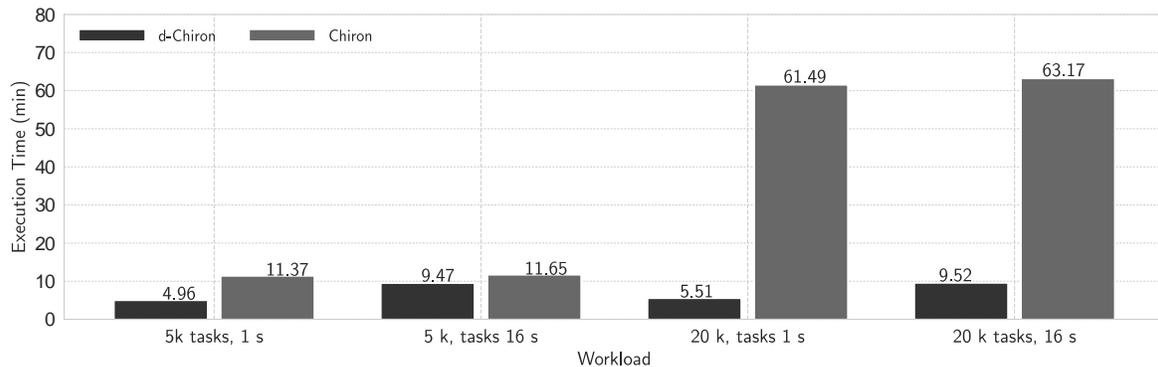

**Figure 14. Comparing d-Chiron with Chiron.**

In the best setting (*i.e.*, large number of short tasks), d-Chiron runs 91% faster than centralized Chiron. By analyzing all results together, we see that the centralized WQ scheduling in Chiron clearly does not scale on this number of cores since it takes approximately the same time executing (a) and (b) and also (c) and (d); hence, almost no performance gain. d-Chiron, on the other hand, completely executes the workload (a) 48% faster than workload (b), and executes (c) 42% faster than (d). Thus, either in large workloads with short-term tasks, or in smaller workloads with long-term tasks, or any combination of these, the in-memory DBMS-driven MTC scheduler outperforms the centralized DBMS-based one. This is because Chiron's centralized design struggles much more than d-Chiron's design following SchalaDB, which allows for direct access from *worker nodes* to the WQ managed by an in-memory DBMS.

## 6. Related Work

We discuss recent workflow execution approaches that take advantage of a DBMS, as SchalaDB proposes. There are DBMS provenance-based systems that support runtime data analysis, which may help workflow adapting [41,45,46,51], but they are disconnected from the workflow execution engine. There are several advantages when the workflow engine integrates workflow execution data and provenance in a unified data management solution. Provenance databases share a lot of information used in task scheduling like tasks, task parameter values, task input data values, dependencies, and task execution time. These information allow for analyses of task execution time, detecting outliers, in addition to a task execution derivation path with its associated input and parameter data [13,26,29].

Several scheduling approaches are moving towards using a DBMS to support their algorithms, like Radical Pilot [31] and Ray [32]. Since scheduling needs high concurrent update transactional support, most solutions adopt MongoDB or similar document DBMSs. Querying MongoDB added with provenance data allows for monitoring the workflow execution. However, similarly to other WMSs that use a DBMS for scheduling, a document-oriented DBMS has limited capabilities for analytical queries, particularly queries that join multiple collections like the ones in Table 2. Next, we analyze WMS related approaches.

EMEWS [22] is a workflow system that uses the Swift/T [14,53] workflow engine, a highly scalable solution, for task scheduling. To keep its scalability, Swift/T engine stores analytical data in log files, which are loaded to a relational DBMS only when the workflow execution finishes, *i.e.*, for *post-mortem* analysis. Still, there are different databases using different DBMSs and data models to manage, and the user does not have access to the data for scheduling, jeopardizing the user steering support. Moreover, e-Science Central [22] is another WMS that uses a graph-oriented DBMS to store execution data at runtime and, after execution, it inserts execution data into a relational DBMS for *post-mortem* analysis.

Pegasus is a scalable WMS that has been used for a large number of real world scientific applications, like LIGO for gravitational wave discovery [10,12,15]. Pegasus uses a DBMS to store workflow data, which is available for the user to do runtime execution monitoring and debugging, but provides a different one through a dashboard for event monitoring, which makes it difficult to do integrated data analyses for user steering.

Stampede [21] is a similar approach to SchalaDB in the sense that it is a DBMS based execution monitoring tool that can be plugged into a WMS. Stampede adopts a centralized DBMS solution and has been evaluated with two different WMSs (Pegasus and Triana) to show its monitoring facilities. However, it is also a solution that does not integrate monitoring to domain or provenance data.

FireWorks is a scalable WMS [25] that also has a DBMS-driven workflow execution engine. FireWorks uses data stored in MongoDB, a document-oriented DBMS, to manage states in queues of the tasks. Querying MongoDB allows for monitoring the workflow execution. However, similarly to other systems that use a DBMS for scheduling, a document-oriented DBMS has limited capabilities for analytical queries that join multiple collections. Moreover, its provenance data representation does not contemplate domain, execution, and provenance data in a same database.

Chiron [34,35] collects provenance data following user's interests and uses a relational DBMS to integrate provenance, execution, and domain data both for the management of parallel execution of tasks and to allow for runtime analytical queries. Chiron is W3C PROC compliant, which helps users in using a uniform provenance data representation for different workflows. However, Chiron employs a centralized approach for the management of data for scheduling of parallel tasks, severely limiting its scalability. Although the DBMS is continuously populated with data for analyses, Chiron implements a traditional master-workers architecture, where the central master node is the only one that is able to access the centralized DBMS. This not only introduces two single points of failure (at the *master* node and at the centralized DBMS), but also, when the number of worker nodes or the WQ is large, Chiron suffers from performance limitations.

We are not aware of any approach that supports user steering and attains high performance workflow execution on large clusters. SchalaDB addresses open challenges described by the scientific community [2,11,16], such as enabling human in the loop for steering and data-aware workload scheduling. When provenance, domain and execution data are all related, integrated in the same database, a user can run analytical queries to help steering [13,49], analyze the results and monitor performance with domain data [43] at runtime. Like Chiron, SchalaDB takes advantage of in-situ processing. According to Ayachit *et al.* [3], *in-situ* processing uses network or shared memory to pass intermediate results. This is the case of the workflow engine working with in-memory distributed DBMS to support workflow data management. Provenance databases register entities (e.g., intermediate results) that are *used by* workflow activities and that are *generated by* them. When the workflow intermediate results are directly passed by the workflow engine to the in-memory provenance database, data movements are avoided, particularly IO with logs. Additionally, the workflow engine can make use of these data for runtime optimizations or adaptive scheduling, without jeopardizing the workflow execution performance.

## 7. Conclusion

In this paper, we proposed SchalaDB, an architecture and a set of techniques based on distributed data management for efficient workflow execution control and for user steering support. Distributed data management techniques such as distributed concurrency control, in-memory data processing, fault tolerance, and others improve the implementation effort in managing scheduling data and other distributed workflow execution data. Using the same DBMS also for workflow provenance data analysis makes the solution more appealing for applications when both high performance and runtime data analyses are essential.

To evaluate its architecture and techniques, we implemented SchalaDB by modifying the management of data for parallel task scheduling and for runtime data analysis of an existing WMS, called Chiron, and we call d-Chiron the newer version with SchalaDB. With SchalaDB we were able to remove all message passing communication related to tasks scheduling, which was present in original Chiron. This approach reduces source code complexity and effort on developing an efficient distributed concurrency control. By doing this, SchalaDB addresses issues inherent to dataflow management, such as data pipelining.

We ran several experiments to validate the scalability of the implementation. We showed that SchalaDB supports scalability very well when executing thousands of long-term tasks (that last at least one minute each), with high scalability in all cases with longer tasks. It is known that tasks in scientific applications require complex calculations usually taking more than a minute each to execute [39]. Finally, we showed that because of SchalaDB, d-Chiron is at least two orders of magnitude faster than original Chiron. Therefore, in addition to attaining high efficiencies on up to 960 cores in an HPC cluster for typical scientific workloads SchalaDB manages provenance, domain, and execution data in the same DBMS providing complex runtime analytical queries, as shown in the experiments. We expect these results to motivate workflow execution engine scientists and engineers to adopt a data-centered solution in their engines. Future work will investigate how SchalaDB can explore different hardware architectures, including GPUs, to manage machine learning workflows.

## Acknowledgments

This work was funded by CNPq, FAPERJ and Inria (HPDaSc associated team). The experiments were carried out using the Grid'5000 testbed from Inria (https://www.grid5000.fr). The authors would also like to thank Pedro Paiva Miranda for his help during the development of d-Chiron.